\documentclass[aps,showpacs,amsmath,preprint,amssymb,12pt]{revtex4}
\usepackage{amsmath,amssymb,bm}
\begin{document}
{~}
\vspace{3cm}

\title{
Black Holes on Eguchi-Hanson Space \\
in Five-Dimensional Einstein-Maxwell Theory
}

\author{Hideki Ishihara\footnote{E-mail: ishihara@sci.osaka-cu.ac.jp}, 
Masashi Kimura\footnote{E-mail: mkimura@sci.osaka-cu.ac.jp},
Ken Matsuno\footnote{E-mail: matsuno@sci.osaka-cu.ac.jp},
and
Shinya Tomizawa\footnote{E-mail: tomizawa@sci.osaka-cu.ac.jp}}

\affiliation{
Department of Mathematics and Physics,
Graduate School of Science, Osaka City University,
3-3-138 Sugimoto, Sumiyoshi, Osaka, 558-8585, Japan}

\begin{abstract}
We construct a pair of black holes on the Eguchi-Hanson space 
as a solution in the five-dimensional Einstein-Maxwell theory.
\end{abstract}
\preprint{OCU-PHYS 249}
\preprint{AP-GR 35}

\date{\today}
\maketitle

In this brief report, we construct a pair of black holes on the 
Eguchi-Hanson space as a solution in the five-dimensional Einstein-Maxwell theory. 
The metric and the gauge potential one-form are given by
\begin{eqnarray}
& & ds^2 = -H^{-2} (r,\tilde\theta) dT^2 + H(r,\tilde\theta) ds^2_{\rm EH}, \label{metric_BH} \\
&& \bm{A} = \pm \frac{\sqrt 3}{2} H ^{-1} (r,\tilde\theta) dT,
\end{eqnarray}
with 
\begin{eqnarray}
&& ds^2_{\rm EH} = \left( 1- \frac{a^4}{r^4}\right) ^{-1} dr^2 
        + \frac{r^2}{4} \left[ \left( 1-\frac{a^4}{r^4} \right) 
          \left( d\tilde\psi+\cos\tilde\theta d\tilde\phi \right)^2
        + d\tilde\theta^2+\sin^2\tilde\theta d\tilde\phi^2 \right],
\label{metric_EH} \\
&& H(r,\tilde\theta) = 1 + \frac{m_1}{r^2 - a^2\cos\tilde\theta}
+ \frac{m_2}{r^2 + a^2\cos\tilde\theta},
\label{harmonics_EH}
\end{eqnarray}
where $a$ and $m_j~(j=1,2)$ are constants, $0\leq \tilde\theta\leq \pi,~ 
0\leq \tilde\phi\leq 2\pi/n,~(n: \mbox{natural number})$ 
and $0\leq \tilde\psi\leq 2\pi$. 

The equation (\ref{metric_EH}) is the metric form of  
the Eguchi-Hanson space\cite{Eguchi}.  
The Eguchi-Hanson space has an S$^2$-bolt at $r=a$, 
where the Killing vector field $\partial/\partial \tilde \psi$ vanishes. 
The function $H(r,\tilde\theta)$ is a harmonics on 
the Eguchi-Hanson space (\ref{metric_EH}).

As is seen later, two black holes are located on 
the north pole ($\tilde\theta=0$) and 
the south pole ($\tilde\theta=\pi$) on the S$^2$-bolt. 
The asymptotic behaviour of the metric (\ref{metric_BH}) 
near the spatial infinity $r\to \infty$ becomes
\begin{equation}
ds^2 \simeq -dT^2 + dr^2 + \frac{r^2}{4}
\left[ d\tilde\theta^2 + \sin^2\tilde\theta d\tilde\psi^2
     +\left( d\tilde\phi + \cos\tilde\theta d\tilde\psi \right)^2 \right].
\end{equation}
Since the $T=$const. surface has the structure of 
lens space $L(2n;1)$, this solution is asymptotically locally flat. 
The Komar mass, $M_{\rm Komar}$ and the total electric charge, $Q$ 
at the spatial infinity are given by  
\begin{equation}
  M_{\rm Komar} = \frac{\sqrt 3}{2}|Q|= \frac{3 \pi (m_1 + m_2)}{8nG},
\end{equation}
where $G$ is the five-dimensional gravitational constant.

In order to clarify the physical properties of the solution, 
we introduce the coordinates as follows
\cite{Prasad},
\begin{eqnarray}
& & R=a\sqrt{\frac{r^4}{a^4}-\sin^2\tilde\theta},\quad 
\tan\theta=\sqrt{1-\frac{a^4}{r^4}}\tan\tilde\theta,\quad 
\phi=\tilde\psi,\quad 
\psi=2\tilde\phi. \notag \\
& &
( 0\leq \theta\leq \pi,~~ 
  0\leq \phi\leq 2\pi,~~
  0\leq \psi\leq 4\pi/n )
\end{eqnarray}
Then, the metric takes the form of 
\begin{equation}
ds^2 = - H^{-2} (R,\theta)dT^2 + H(R,\theta) ds^2_{\rm EH},
\end{equation}
with
\begin{eqnarray}
& & ds^2_{\rm EH} = V ^{-1} (R,\theta) 
               \left[ dR^2 + R^2 \left( d\theta^2 + \sin^2\theta d\phi^2 \right) \right]
               + V(R,\theta) \left( \frac{a}{8} d\psi+\omega_\phi d\phi \right)^2,
\label{metric_GH}
\\
& & H(R,\theta) = 1 + \frac{m_1/a}{|{\bm R}-{\bm R}_1|}
             + \frac{m_2/a}{|{\bm R}-{\bm R}_2|},\\
& & V^{-1} (R,\theta) = \frac{a/8}{|{\bm R}-{\bm R}_1|}
            +\frac{a/8}{|{\bm R}-{\bm R}_2|},\\ 
& &\omega_{\phi}(R, \theta)
= \frac{a}{8} 
        \left(
  \frac{R \cos\theta - a}{\sqrt{R^2 + a^2 - 2 a R \cos\theta}}
+ \frac{R \cos\theta + a}{\sqrt{R^2 + a^2 + 2 a R \cos\theta}}
\right),
\end{eqnarray}
where ${\bm R}=(x,y,z)$ is the position vector on 
the three-dimensional Euclid space and ${\bm R}_1=(0,0,a)$, 
${\bm R}_2=(0,0,-a)$. 
The metric (\ref{metric_GH}) is the  
Gibbons-Hawking two-center form of the Eguchi-Hanson 
space\cite{GH,Prasad}. 
It is manifest in the coordinate that the space has 
two nut singularities at $\bm R= \bm R_j$ 
where the Killing vector field $\partial /\partial \psi$ vanishes.

The function $H(R,\theta)$ is the harmonics given by (\ref{harmonics_EH})
on the Eguchi-Hanson space in the Gibbons-Hawking coordinates (\ref{metric_GH}).
The harmonics $H(R,\theta)$ converts nut singularities 
on the Eguchi-Hanson space to regular hypersurfaces in the total 
spacetime.  
Since each hypersurface ${\bm R}={\bm R}_j$ 
becomes a Killing horizon with respect to the Killing vector field $\partial/\partial T$, 
and each three-dimensional section of them with $T={\rm const.}$ 
has finite area, then the hypersurfaces ${\bm R}={\bm R}_j$ are 
event horizons. 

Since
the Kretschmann invariant $R_{\mu\nu\rho\lambda}R^{\mu\nu\rho\lambda}$ 
has a finite value on each horizon, we see that 
the geometry on the horizons is regular. Even if one of $m_j$ (for an example $m_2$) vanishes, 
which corresponds to a single black hole and a naked nut charge 
with the value $a/8$, the horizon is regular. The spacetime is regular in the case of $n=1$ but 
 it has a nut singularity at ${\bm R}={\bm R}_2$ in the case of $n \ge 2$.

The induced metric on the spatial cross section of the $j$-th horizon is given by
\begin{equation}
ds^2_{\rm Horizon}
= \frac{m_j}{8} \left[ d\theta^2+\sin^2\theta d\phi^2 
     + \left( d\psi + \cos\theta d\phi \right) ^2 \right], ~~(0\leq \psi \leq 4\pi/n)
\end{equation}
which is the lens space $L(n;1)$.
The geometry
near horizons of this solution  
is similar to the multi-black hole solutions 
on the Gibbons-Hawking multi-instanton space\cite{IMKT}, 
but the asymptotic structures are different. 
Although both are asymptotically locally flat,  
the former is isotropic in four spatial dimensions
while the latter has a compact dimension 
as same as Kaluza-Klein black holes discussed in Ref.\cite{IMKT,IM}. 

\section*{Acknowledgements}
We thank K. Nakao and Y. Yasui for useful discussions. 
This work is supported by the Grant-in-Aid
for Scientific Research No.14540275.

\end{document}